\documentstyle[12pt,epsfig,graphics]{article}
\newcommand{\be}{\begin{equation}}
\newcommand{\ee}{\end{equation}}
\newcommand{\ba}{\begin{eqnarray}}
\newcommand{\ea}{\end{eqnarray}}
\newcommand{\lb}{\label}
\newcommand{\bb}{\bibitem}
\newcommand{\half}{\frac{1}{2}}

\newcommand{\nn}{\nonumber}
\begin{document}

\begin{center}
{\LARGE Spacetime Defects: von K\'arm\'an vortex street like configurations}

\vspace{1cm}

 {\large 
  Patricio S. Letelier\footnote{e-mail: 
letelier@ime.unicamp.br} }
\vspace{1ex} 
 
  Departamento de Matem\'atica Aplicada-IMECC\\
Universidade Estadual de Campinas\\
13081-970 Campinas. S.P., Brazil

\end{center}
\vspace{4ex} 

\centerline{ \small\bf Abstract}
A special arrangement of spinning strings with dislocations similar to a 
von K\'arm\'an vortex street  is studied. We numerically  solve the geodesic
equations for the special case of a test particle  moving along two
infinite rows of pure dislocations and also discuss the case of pure
spinning defects.

\hspace{0.3cm} 

\hspace{1cm} 
\baselineskip 0.7cm

Conical singularities or spacetime defects are  characterized 
by Riemann-Christoffel curvature tensor, or Cartan torsion,  or
 both different from zero only on the subspace (event, world line, world 
sheet, or world tube) that describes the evolution of the defect (texture, 
monopole, string, or membrane). In other words we have
that the curvature, the torsion, or both, are proportional
 to distributions with support on the defect. 
Spacetimes with conical  singularities 
of different types has been studied recently in a variety of contexts, e.g.
spinning  strings with cosmic dislocations \cite{gallet}\cite{lchi}, 
pure spacetime dislocations \cite{tod}-- \cite{loop}, also in low
 dimensional gravity \cite{koh}. For the discussion of a great variety 
of defects see Ref. \cite{lew}.

Some quantum aspects related to line defects has been considered by
 several authors:
The spectra of a quantum particle in the presence of conical
 singularities that physically corresponds to a cosmic strings, 
 a  screw dislocation,
and a rotating string \cite{bez1997}. The harmonic interaction
of a particle in the presence of  defects \cite{furtado2000}.
A Berry quantum phase for particles transported along a
 defect \cite{deassis2000}. 
The relation with the chiral anomaly is Ashtekar approach to quantum gravity 
\cite{mielke}.
Classical aspects has been also recently considered: The gravitational
 energy of 
the conical defects
\cite{maluf}. Geodesic motions around line defects \cite{depadua}. The
 magnetic and electric self-forces
on a straight wire induced by a topological line defect carrying singular
 torsion \cite{carvalho2000}.

The purpose of this note is to study a particular infinite 
superposition of line defects that can be described as
  a von K\'arm\'an vortex street \cite{milne-t} kind
of arrangement of spinning  cosmic strings with dislocations. This
 arrangement of strings
is depicted  in Fig. 1. We have two rows of an infinite number of  strings,
 the coordinate distance between the rows is $2b$ and the strings in a row are
 apart by a distance $2a$. One row is displaced  in an amount $a$ with respect
 to the other.
All the strings in the same  row have equal  spins and equal dislocations
 (the spins and the dislocations are not necessarily equal). The strings in the 
two  rows  have opposite spins and dislocations.

First, let us summarize the main relations associated to a
 spinning string with dislocation. The associated spacetime is
 \cite{lchi},
\be
ds^2= (\omega^t)^2 - (\omega^z)^2  - (\omega^x)^2  - (\omega^y)^2, \lb{m1}
\ee
with
\ba 
&&\omega^t=dt-\partial_y W d x  +\partial_x W d y ,  \lb{om0}\\ 
&&\omega^z=dz-\partial_y U d x+\partial_x U d y, \lb{om3}\\
&&\omega^x =e^{-2V} dx,\;\; \omega^y =e^{-2V} dy, \lb{om12}
\ea
where the the functions: $U, W$ and $V$ are 
\be  
W=4\sigma f_W, \;\; U=4\kappa f_U, \;\; V=2\lambda f_V, \lb{wuv}
\ee
and  for the case of a single defect all the three functions, $f_W, \; f_U$ and
$f_V$ are equal
\be
f=\ln r, \;\; (r\equiv+\sqrt{x^2+y^2}) . \lb{f1}
\ee
In the context of the Riemann-Cartan geometry\cite{gockeler} this spacetime
has  curvature and torsion that are proportional to Dirac distributions 
with support on the
line $x=y=0$. We have that the tetradical 
components of the curvature tensor and the torsion reduce to:
\ba
&& R_{xyxy}= 2e^{4V}(\partial_{xx}+\partial_{yy})V=
8\pi\lambda \delta(x)\delta(y)/\sqrt{-g}, \lb{r1}\\
&&S_{txy}=(\partial_{xx}+\partial_{yy})W=8\pi \sigma
 \delta(x)\delta(y), \lb{t1a}\\
&&S_{zxy}=(\partial_{xx}+\partial_{yy})U=
8\pi \kappa \delta(x)\delta(y). \lb{t1b}
\ea
>From  Einstein-Cartan equations one finds that this metric represents a 
single string locate along the $z$-axis with
 linear density, $\lambda$ \cite{lchi}. The string has as 
 equation of state: linear density equal to  tension, i.e., the
 equation of state for usual cosmic strings.  The string is spinning with an 
``angular velocity'' $\sigma$ and has a dislocation given by $\kappa$. In
 this case, the  analogue of the Burgers vector of dislocation  has a 
single component along the $z$-axis, $2\kappa/\pi$.

Note that we  have three independent structures: i) A disclination or
cosmic string, ii) A  ``time dislocations'' or  spin,  and ii) A
  space dislocation or cosmic dislocation. Each one of these structures gives
us a singularity along the $z$-axis (line singularity).
Thus, when these three structures are one in top of the other we have the a 
spinning string with cosmic dislocation. But we can also  have:
a) A string with dislocation only, $\sigma=0$, b) A spinning string,
 $\kappa=0$, c) A spinning cosmic dislocation,  $\lambda=0$,
d) A pure  time dislocation or spinning defect, $\lambda=\kappa=0$, e) A pure
 cosmic dislocation, $\lambda=\sigma=0$, and  f) A usual cosmic string, 
$\kappa=\sigma =0$. The gravitational energy of these defects were considered in \cite{maluf} and the spacetime  analogue of other  Volterra distortions 
in \cite{punti}.

Now we shall consider single row of strings form by identical 
 strings parallel to the 
$z$-axis that cross the $x$-axis, at: $ \cdots -4a, -2a, 0, +2a, +4a, \cdots$. 
The function $f_V$ in this case is
\be
f_V=\half\sum^{\infty}_{n=-\infty}\ln[(x-2na)^2+y^2]. \lb{fs0}
\ee
This function gives right curvature, since
\be
(\partial_{xx}+\partial_{yy})f_V=2\pi\sum^{\infty}_{n=
-\infty}\delta(x-2na)\delta(y).
\lb{deltas}
\ee
Introducing the complex variable, $\zeta=x+iy$, we have,

\ba
2f_V&=&\sum^{\infty}_{n=-\infty}\ln[(\zeta-2an)(\bar\zeta -2an)], \lb{fs1}\\
& =& \ln(\zeta\bar\zeta)+\sum^{\infty}_{n=1}\ln[(\zeta^2-4a^2n^2)
(\bar{\zeta}^2-4a^2n^2)], \lb{fs2}\\
&  =&2\widehat{f}_V+4\sum^{\infty}_{n=1}\ln(2an) \lb{fs3}
\ea
with
\be
2\widehat{f}_V=\ln\{\zeta\prod_{n=1}^{\infty}[1-(\zeta/2an)^2]
\bar\zeta \prod_{m=1}^{\infty}[1-(\bar\zeta/2am)^2]\} \lb{fsf}
\ee
Since the singular structure of (\ref{deltas}) does
 not change if we add a constant to $f_V$ we have that  (\ref{fs3})
and (\ref{fsf}) represent the same type of defects. So we shall use 
this last function to represent the infinite row of strings.  By using
 the identity \cite{ww},
\be
\sin x = x \prod_{n=1}^{\infty}[1-(x/n\pi)^2],
\ee
$\widehat{f}_V$ can be cast  in the simple form,
\ba
\widehat{f}_V(x,y) &=& \ln|\sin[\pi(x+iy)/2a]|, \lb{fs4}\\
 &=&\half\ln[\cosh^2(\pi y/2a)-\cos^2(\pi x/2a)]. \lb{fs5}
\ea

Thus the functions $V=\lambda\widehat{f}_V(x,y)$ 
 and $W=U=0$  are  the metric potentials that describe an infinite row
 of cosmic strings. With $\widehat{f}_V(x,y-b)$ we have the upper row of 
Fig. 1 with no spin and dislocation.  To put  spin and dislocation to each string we do
  $W=\sigma \widehat{f}_V(x,y-b)$ and  $U=\kappa\sigma \widehat{f}_V(x,y-b)$.
 The other row 
is represented by the functions
$V=\lambda\widehat{f}_V(x+a,y+b),\; $ $W=-\sigma\widehat{f}_V(x+a,y+b)\;$
 and $U=-\kappa\widehat{f}_V(x+a,y+b)$
(the spins, as well as, the dislocations of the defects of the two 
rows are opposite). Now the metric potentials for the
whole von K\'arm\'an vortex street type of distribution
 presented in Fig. 1 are:  $W=\sigma [\widehat{f}_V(x,y-b)-  
\widehat{f}_V(x+a,y+b)]\;$,  $U=\kappa [\widehat{f}_V(x,y-b)-  
\widehat{f}_V(x+a,y+b)]\;$, and   $V=\lambda [\widehat{f}_V(x,y-b)+  
\widehat{f}_V(x+a,y+b)]\;$, i.e.,
\ba
&&W=2\sigma\ln\frac{\cosh^2[\pi( y-b)/2a]-\cos^2(\pi x/2a)}
{\cosh^2[\pi( y+b)/2a]-\sin^2(\pi x/2a)}, \lb{wvk}\\
&&U=2\kappa\ln\frac{\cosh^2[\pi( y-b)/2a]-\cos^2(\pi x/2a)}
{\cosh^2[\pi( y+b)/2a]-\sin^2(\pi x/2a)}, \lb{uvk}\\
&&V=\lambda\ln\{(\cosh^2[\pi( y-b)/2a]-\cos^2(\pi x/2a))\times \nn\\
&& \hspace{1.5cm}(\cosh^2[\pi( y+b)/2a]-\sin^2(\pi x/2a))\} \lb{vvk}
\ea

In order to better understand this metric we shall study the motion of test
particles  traveling between the two rows of strings. Since the torsion
is a distribution with support only on the defect we have that, in this case,
 geodesics and  autoparallels are the same. The geodesic equations for
 the metric (\ref{m1})-(\ref{om12}) reduce to:
\ba 
&&\dot t- \dot x\partial_y W  +  \dot y \partial_x W =K_1 ,  \lb{geo1}\\ 
&&\dot z- \dot x\partial_y U +  \dot y\partial_x U=K_2 , \lb{geo2}\\
&&\ddot x -2(\dot{x}^2-\dot{y}^2)\partial_x V -4\dot x \dot y\partial_yV  =0, \lb{geo3}\\ 
&&\ddot y -2(\dot{x}^2-\dot{y}^2)\partial_y V-4 \dot x \dot y\partial_x V =0, \lb{geo4} 
\ea
where $K_1$ and $K_2$ are constants of integration.
 From the metric (\ref{m1})-(\ref{om12}) we obtain the relation,
\be
(\dot{x}^2+\dot{y}^2)e^{-4V}=-1+K_{1}^2 -K_{2}^2. \lb{cm}
\ee
We note that the motion in the plane $(x,y)$ is independent of the
 variable $z$ and is determined mainly by the string part. 
The motion in this plane determines the motion along $z$. This
 last motion is due solely to the  dislocation. Since this part will
 represent the major depart from the usual cosmic strings we shall first
  analyze the case:
$\lambda=\sigma=0$, i.e., an arrangement of pure dislocations. Now the
 geodesic
 equations are: (\ref{geo2}),  $\dot t=K_1$, and $\ddot x =\ddot y =0$.
The last two equations tell us that the motion is confined to a plane parallel
to the $z$-axis. We integrate these equation and choose the constant of
 integrations in such a way that the particle be confined between
 the two rows  of dislocations: $y=y_0$ with $-b<y_0<b$  and $x=v_{x}s+x_0$.
The constraint (\ref{cm}) now reads,
\be
K_1^2=1+K_2^2+v_x^2 . \lb{cm2}
\ee
Thus, it remains  only one equation to solve, Eq. (\ref{geo2}), that we 
shall study numerically. 
In Fig. 2 we present the geodesic motion of a test particle in the
 gravitational field of a von K\'arm\'an
  street type of configuration constructed with cosmic dislocations
 for different values of the constants: $y_0=-1.8$ 
(lower curve near the origin), $y_0=0$ (middle curve) and $y_0=1.8$ (upper curve); $x_0=0$ for the three cases. The value of the parameters are $a=1, \; b=2, \; \kappa=-0.45$.
We take for the integration constants: $v_x=0.8$ and $K_2=0$ [note that $K_1$
is determined by Eq. (\ref{cm2})]. The three curves intersect in points 
that have the same  coordinate $x$ as  the dislocations.

In summary, with no cosmic  dislocations  we have a motion
 on the plane $z=0$ parallel
 to the $x$-axis.  The presence of the pure cosmic dislocations makes 
the particle
 leave this plane and acquire  velocity along the $z$ direction.
Depending on the sign,  the dislocations may accelerate or brake the 
motion of the particle in the $z$-direction. We have a very symmetric 
arrangement of dislocations that is reflected in the motion of the particles.

Finally, let us turn our attention to the case of pure spinning defects,
 in this case we  have $\kappa=\lambda =0$. Then the particles move always 
in a strait line.  If we take $\sigma=-0.45$ and the same
 values for  the rest of the constants as in the precedent case, we find that
 the test particles do not leave the plane $(x,y)$ and 
move  parallel to the $x$-axis. The time
coordinate is ``accelerated'' and ``braked'' as the $z$-coordinate we
 have the same behaviour shown in Fig. 2 with the $z$-axis changed by the 
coordinate $t$. Again the symmetry of the   arrangement of pure  spinning defects  
 is reflected in the motion of the test particles.

\vspace{0.5cm}
\noindent
Acknowledgements.--
I want to thank FAPESP and CNPQ for financial support and also Prof.
 Boisseau
for several discussions concerning strings and other extended objects.



\newpage
\begin{center}
\begin{figure}
\epsfig{width=3in,height=5in,angle=-90,file=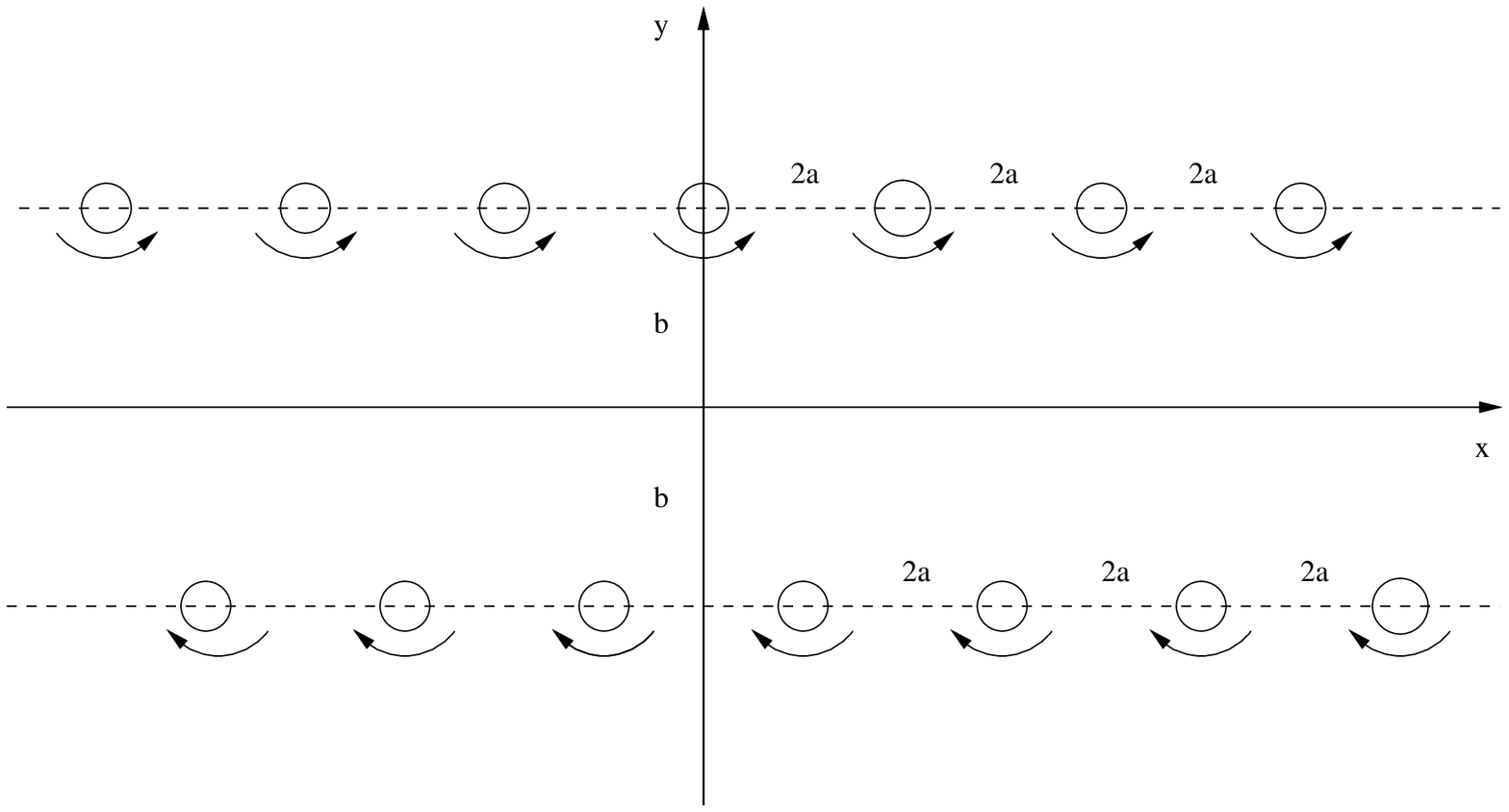}
 
\caption{A von K\'arm\'an vortex street
 like distribution of spinning cosmic strings with dislocations 
is depicted. All the strings in the same  row have equal  spins
 and equal dislocations
 (the spins and the dislocations are not necessarily equal). The strings in the 
two  rows  have opposite spins and dislocations.  }
\end{figure}
\end{center}

\newpage
\begin{center}
\begin{figure}
\epsfig{width=3in,height=5in,angle=-90,file=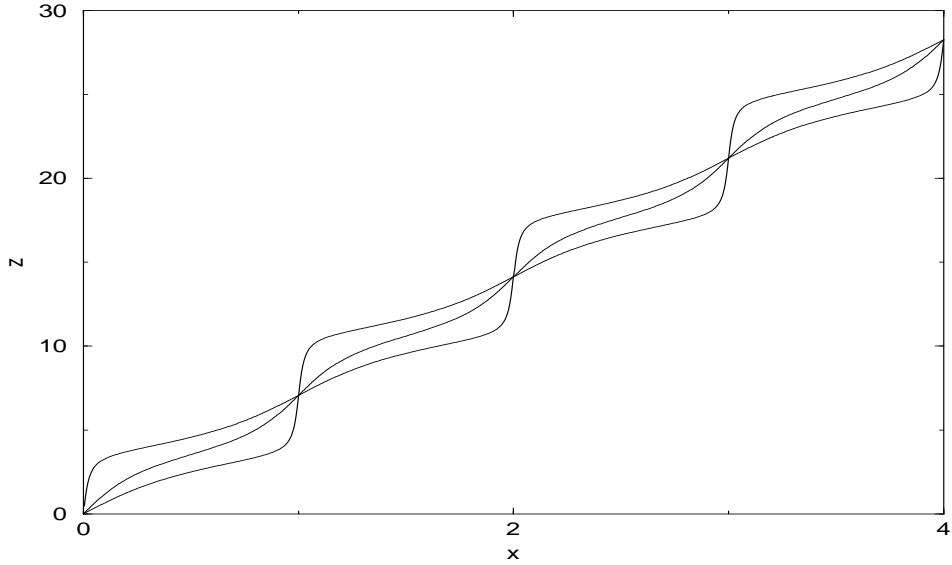}
 \caption{ Geodesic motions of a test particle in 
a von K\'arm\'an street type of distribution of 
cosmic dislocations for different values of the constants: $y_0=-1.8$ 
(lower curve near the origin), $y_0=0$ (middle curve) and $y_0=1.8$ (upper curve); $x_0=0$ for the three cases. The value of the parameters are $a=1, \; b=2, \; \kappa=-0.45$.
We take for the integration constants: $v_x=0.8$ and $K_2=0$.
   }
\end{figure}
\end{center}

\end{document}